\def\year{2022}\relax
\documentclass[letterpaper]{article} 
\usepackage{aaai22}  
\usepackage{times}  
\usepackage{helvet}  
\usepackage{courier}  
\usepackage[hyphens]{url}  
\usepackage{graphicx} 
\usepackage{natbib}  
\usepackage{caption} 
\DeclareCaptionStyle{ruled}{labelfont=normalfont,labelsep=colon,strut=off} 
\usepackage{algorithm}
\usepackage{algorithmic}
\usepackage{microtype}
\usepackage{multirow}
\usepackage{amssymb}
\usepackage{amsmath}
\usepackage{booktabs}
\usepackage{rotating}
\usepackage{xcolor}

\usepackage{newfloat}
\usepackage{listings}
\floatstyle{ruled}
\newfloat{listing}{tb}{lst}{}
\floatname{listing}{Listing}
\title{Radiology Report Generation with a Learned Knowledge Base \\ and Multi-modal Alignment}
\author{
    Shuxin Yang\textsuperscript{{\rm 1},{\rm 5}},
    Xian Wu\textsuperscript{\rm 3},
    Shen Ge\textsuperscript{\rm 3},
    S.~Kevin~Zhou\textsuperscript{{\rm 1},{\rm 2}},
    Li Xiao\textsuperscript{{\rm 1},{\rm 4}}
}
\affiliations{

    \textsuperscript{\rm 1}The Key Lab of Intelligent Information Processing of Chinese Academy of Sciences (CAS), Institute of Computing Technology, CAS, Beijing, 100190, China \\
    \textsuperscript{\rm 2}School of Biomedical Engineering \& Suzhou Institute for Advanced Research Center for Medical Imaging, Robotics, and Analytic Computing \& LEarning (MIRACLE) University of Science and Technology of China, Suzhou 215123, China \\
    \textsuperscript{\rm 3}Tencent Medical AI Lab, Beijing, 100094, China \\
    \textsuperscript{\rm 4}University of Chinese Academy of Sciences, Beijing, 100049, China
}

\usepackage{bibentry}

\begin{document}

\maketitle

\begin{abstract}
In clinics, a radiology report is crucial for guiding a patient's treatment. However, writing radiology reports is a heavy burden for radiologists. To this end, we present an automatic, multi-modal approach for report generation from a chest x-ray. Our approach, motivated by the observation that the descriptions in radiology reports are highly correlated with specific information of the x-ray images, features two distinct modules: (i) \textbf{Learned knowledge base}: To absorb the knowledge embedded in the radiology reports, we build a knowledge base that can automatically distill and restore medical knowledge from textual embedding without manual labor; (ii) \textbf{Multi-modal alignment}: to promote the semantic alignment among reports, disease labels, and images, we explicitly utilize textual embedding to guide the learning of the visual feature space. We evaluate the performance of the proposed model using metrics from both natural language generation and clinic efficacy on the public IU-Xray and MIMIC-CXR datasets. Our ablation study shows that each module contributes to improving the quality of generated reports. Furthermore, the assistance of both modules, our approach outperforms state-of-the-art methods over almost all the metrics.
\end{abstract}


\section{Introduction}
The radiology report is crucial for assisting clinic decision making~\citep{zhou2019handbook}.It describes some observations on images such as diseases' degree, size, and location. However writing radiology report is both time-consuming and tedious for radiologists ~\citep{bruno2015understanding}. Therefore, With the advance of deep learning technologies, automatic radiology report generation has attracted growing research interests. 

Existing radiology report generation methods usually follow the practice of image captioning models~\citep{xu2015show,lu2017knowing,anderson2018bottom}. For example, \citep{Jing2018,Yuan2019} employ the encoder-decoder architecture and propose the hierarchical generator as well as the attention mechanism to generate long reports. However, the radiology report generation task is different from the image captioning task. In image captioning, the model is required to cover all details of the input image. In contrast, for radiology report generation, the model is required to focus on the abnormal regions and infer potential diseases. Therefore, to generate an accurate radiology report, the model needs to identify the abnormal regions and provide proper descriptions of specific information of diseases. To this end, the medical knowledge needs to be included in modeling.  

Recently, some works attempt to integrate medical knowledge in modeling: the Meets Knowledge Graph(MKG)~\cite{Zhang2020when} and Poster and Prior Knowledge Exploring and Distilling(PPKED)~\cite{Liu2021Exploring} incorporate manual pre-constructed knowledge graphs to enhance the generation, the Hybrid Retrieval-Generation Reinforced(HRGR)~\cite{Li2018Hybrid} builds a template database based on prior knowledge by manually filtering a set of sentences in the training corpus. These methods achieve improved performance over image captioning models. However, these models need extra labour to build the knowledge graph or template database in advance, which is still laborious. In addition, when applying these models to images of other diseases, the knowledge graph or template database needs to be rebuilt, making it difficult to migrate these methods to other datasets.

In this paper, we propose a knowledge base updating mechanism to learn and store medical knowledge automatically. Firstly, we initialize a memory module as a knowledge base and use CNN/BERT model to extract visual features and textual embeddings from the input images and their corresponding reference reports, respectively. Secondly, the knowledge base is updated by the report embeddings during the training phase. At the end of the training, we fix the knowledge base as the model's parameter and use it for inference. To retrieve the relevant knowledge of the input image, we propose a visual-knowledge attention module that queries knowledge base with visual features. Finally, we combine acquired knowledge with visual features to generate radiology reports.

Since the critical clinical information usually comes from descriptions of abnormalities, where such sentences are rare and diverse in radiology datasets, we need to enable the knowledge base to focus on the knowledge of abnormalities. To this end, we propose a multi-modal alignment mechanism. It consists of visual-textual alignment and visual-label alignment. The intuition is that the reports and disease labels describe the same observations on the images, thus the semantic features among images, reports, and disease labels are likely to be consistent. Specifically, we adapt the triplet margin loss~\citep{Balntas2016} to align the visual features and textual embeddings, as well as the binary cross-entropy loss to align the visual features and disease labels. Guided by the proposed multi-modal alignment, the proposed knowledge base can store the knowledge of abnormalities and generate more accurate descriptions of abnormalities.

We evaluate our proposed methods on the publicly accessible IU-Xray and MIMIC-CXR datasets. Besides natural language generation (NLG) metrics, we adopt clinical efficacy (CE) metrics to analyze the quality of generated reports from the perspective of clinics. The results show that the proposed method achieves state-of-the-art performance on both NLG and CE metrics. It also indicates that the radiology report generation benefits from the multi-modal alignment mechanism and the learned knowledge base, avoiding laborious manual construction of the knowledge graph or template database. Furthermore, the proposed methods boost the quality of generated reports in both natural language and clinical correctness.

The main contributions are as follows:
\begin{itemize}
    \item We propose a novel radiology generation framework with a learned knowledge base, which could learn and store medical knowledge automatically during training by a novel knowledge updating mechanism without any manual labor.
    \item We propose a multi-modal alignment mechanism that promotes the semantic alignment among images, reports, and disease labels to guide the learning of visual features.
    \item Experiments demonstrate that the proposed components achieve consistent performance improvements. Furthermore, our model achieves state-of-the-art performance over almost all metrics for both of the public IU-Xray and MIMIC-CXR datasets. 
\end{itemize} 

\section{Related Work}

With the advancement of computer vision and natural language processing, many works exploit to combine radiology images and free-text for automatically generating reports to assist radiologists in the clinic~\cite{zhou2021review}. Inspired by image captioning, \cite{Shin2016} adopts the CNN-RNN framework to describe the detected diseases based on visual features on a chest x-ray dataset. The work is restricted to the categories of predefined diseases. The Co-Att~\cite{Jing2018} and \cite{Xue2018,Yuan2019} propose different attention mechanisms and hierarchical LSTM to generate radiology reports. However, most reports generated by such works tend to describe global observations, which indicates that such methods have difficulties capturing subtle changes in the image. Our proposed method is similar to the TieNet~\cite{Wang2018} which proposes an attention encoded text embedding and a saliency weighted global average pooling to boost the image classification and report generation. However, TieNet uses the report embeddings as a part of LSTM's input for training which is not available in the inference stage, resulting in embedding bias in the report generation.

Other types of work explore injecting extra prior knowledge into the generation model to improve the quality of the generated radiology reports. Following the writing practice of radiologists, HRGR~\cite{Li2018Hybrid} compiles a manually extracted template database to generate radiology reports by reinforcement learning. The Knowledge-driven Encode,Retrieve Paraphrase(KERP)~\cite{Li2019Knowledge}, MKG~\cite{Zhang2020when} and PPKED~\cite{Liu2021Exploring} propose to combine pre-constructed knowledge graph for radiology report generation. Although these methods achieved remarkable results, building the template database and knowledge graph is still laborious, making it hard to transfer those approaches directly to other datasets. This paper proposes a knowledge-enhancing generator to build the knowledge base without manual labor automatically. 

As revealed by recent studies~\cite{Graves2016Hybrid, Wu2018,ChenImp2020}, the memory mechanism can provide prior knowledge to boost the generation model. The R2Gen~\cite{Chen2020} proposes a relational memory in the decoder to learn the order of words or sentences. The major difference between R2Gen and the proposed model is
how the memory is built. R2Gen builds memory from the already generated words of a sentence. While our works build memory from the entire medical report. In this manner, R2Gen can only mine the mutual relation of words within a sentence while our work is able to mine the mutual relation within a report. As a result, our work can identify some long distance relations which can benefit medical report generation.

The key-value memory networks(KVMN)~\cite{Miller2016KVMN} proposes a key-value structured memory that automatically encodes prior knowledge into the memory for question and answer tasks (QA). The knowledge base in this paper is different from the QA and NLG tasks. The proposed knowledge base update mechanism only learns the knowledge from the training data and keeps it fixed during the inference stage. In contrast, The QA and NLG tasks typically require additional knowledge during the training and inference stages. Specifically, our work is different from KVMN according to three aspects. Firstly, KVMN, which deals with the QA task, is a classification model and thus cannot be directly applied to the report generation task. Secondly, the knowledge source of KVMN is pre-defined by a manually developed database, while our knowledge base is learned from scratch during training without manual labor. Finally, the multi-head updating mechanism proposed in our work has more representation ability than the KVMN model.

\begin{figure*}[t]
  \centering
  \includegraphics[width=\linewidth]{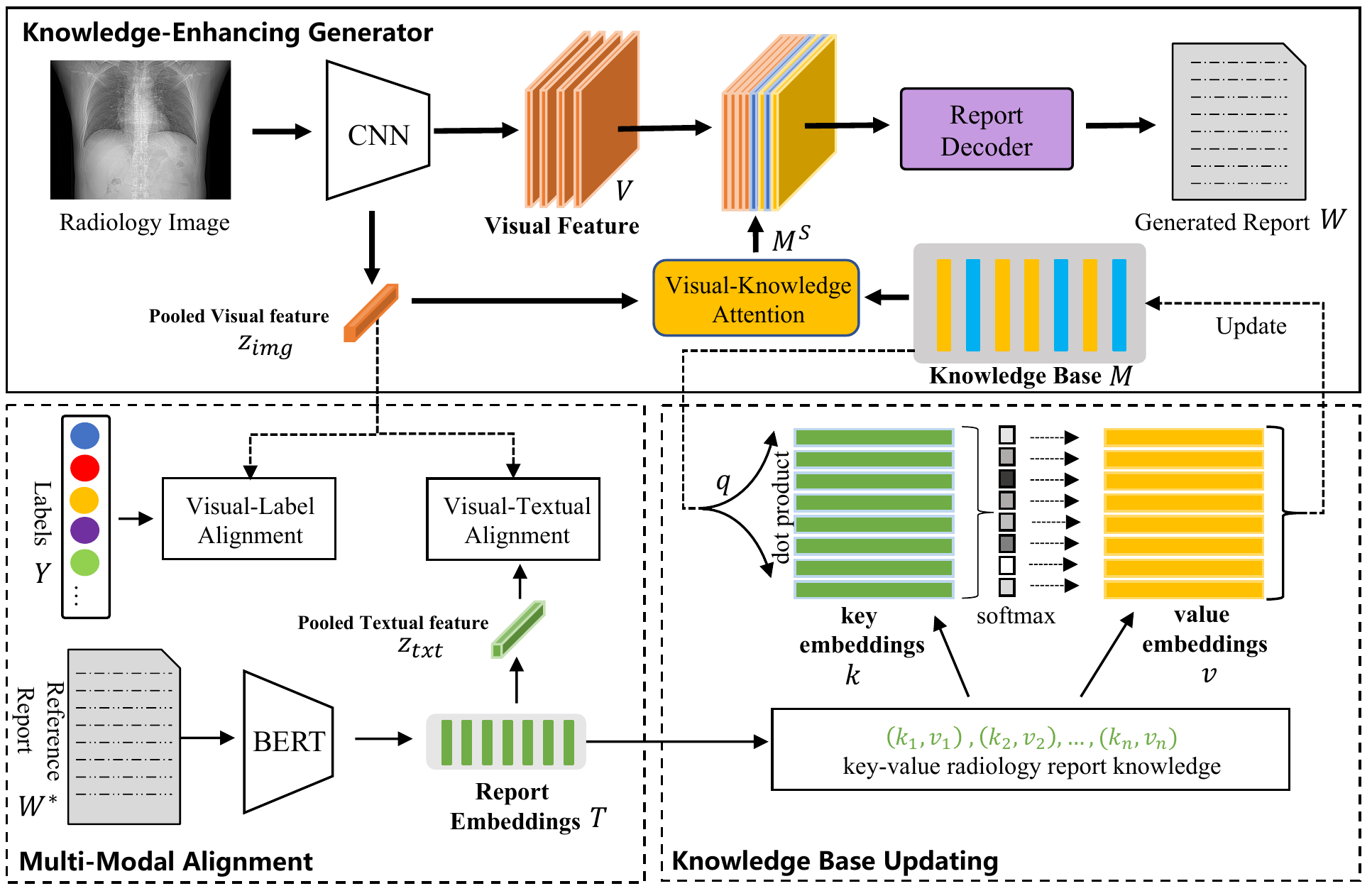}
  \caption{The architecture of the proposed model includes a knowledge-enhancing generator, a knowledge-updating module, and a multi-modal alignment module. The modules contained in the dashed line boxes are only used in the training stage. The knowledge base updating module and the multi-modal alignment module work together to learn the knowledge base and optimize multi-modal semantic alignment. The knowledge base is fixed during inference, and the knowledge-enhancing generator extracts related knowledge from the knowledge base to generate radiology reports.}
  \label{fig:model}
\end{figure*}

\section{Method}
In this section, we introduce the proposed method. Firstly, we provide the notation and formulate the radiology report generation task; Secondly, we present the framework of our method; Thirdly, we describe the proposed knowledge base updating and multi-modal alignment mechanism in detail.

\subsection{Notation and Problem Formulation}
\label{sec:formulation}
Let $Img$ denotes a radiology image, $\mathbf{Y}=\{y_1,\ldots,y_{N_L}|y_i=0/1\}$ denotes the class label of this image, $\mathbf{W^*}=\{w^*_1, w^*_2,\ldots, w^*_{N_R}|w^*_i \in \mathbb{V}\}$ denotes the reference (ground truth) report, $\mathbf{W}=\{w_1, w_2,\ldots, w_{N_W}|w_i \in \mathbb{V}\}$ denotes the report generated by the model. Here $w^*_i$ and $w_i$ refer to the index of word in vocabulary $\mathbb{V}$, $N_L$ is the number of labels, $N_R$ and $N_W$ are the length of the reference and generated reports. In the training stage of the proposed model, we need $Img$, $\mathbf{Y}$ and $\mathbf{W^*}$; In the inference stage, we only need $Img$ to generate $\mathbf{W}$.

\subsection{Framework}
\label{sec:overview}
Figure \ref{fig:model} displays the framework of the proposed model: a knowledge-enhancing generator, a knowledge-updating module, and a multi-modal alignment module. During training, the knowledge-updating module and the multi-modal alignment module work together to store learned knowledge into the knowledge base; During inference, the knowledge base is fixed, and the knowledge-enhancing generator extracts related knowledge from the knowledge base to generate radiology reports.

Firstly, we extract visual features from the input radiology image. Following previous works~\cite{Jing2018,Wang2018,Chen2020}, the convolution neural network (CNN) is employed as our visual encoder, and the visual features are extracted from the last convolution layer:
\begin{align}
    \mathbf{V} &= \text{CNN}(Img)W^E, \label{eqn:visualfeature}
\end{align}
where $W^E$ are learnable parameters for affine transformation, $\mathbf{V} \in \mathbb{R}^{K \times D}$ represents the extracted visual features, where $K$ and $D$ denote the number of the visual features and the dimension of each feature, respectively. 

Next, we acquire the aggregated visual features by average pooling:
\begin{align}
    \mathbf{z}_{img} = \text{AvgPooling}(\mathbf{V}), \label{eq:zimg}
\end{align}
where $\mathbf{z}_{img}$ has the dimension of $D$ and $\text{AvgPooling}$ refers to average pooling function. 

During training, we also extract report embeddings from reference reports to update the knowledge base and guide the learning of visual features. Similar to BERT model~\cite{Devlin2019}, we employ the encoder of Transformer~\cite{Vaswani2017} as our report encoder to extract report embeddings. The report embeddings are acquired from the hidden states of the last layer:
\begin{equation}
    \mathbf{T} = \text{BERT}(\mathbf{W^*}), \label{eqn:textfeature}
\end{equation}
where $\mathbf{W^*}$ denotes the reference reports written by radiologists, $\text{BERT}(\cdot)$ refers to the report encoder, and $ \mathbf{T} \in \mathbb{R}^{N_R \times D}$ is extracted textual embedding.

In BERT, the first token (\textit{CLS}) of a sentence is regarded as the aggregate representation of the entire sequence. Based on the report embedding acquired in Eq.(\ref{eqn:textfeature}), the aggregated textual feature of the report $\mathbf{z}_{txt}$ is acquired by:
\begin{align}
    \mathbf{z}_{txt} = \mathbf{T}_{[CLS]}W^Z + b^Z, \label{eqn:ztxt}
\end{align}
where $\mathbf{T}_{[CLS]}$ denotes the embedding of \textit{CLS} in report embeddings $\mathbf{T}$, $W^Z$ is a learnable affine transformation, and $b^Z$ is a bias. Thus, $\mathbf{z}_{txt}$ has the dimension of $D$.

\begin{figure}[htbp]
  \centering
  \includegraphics[width=0.87\linewidth]{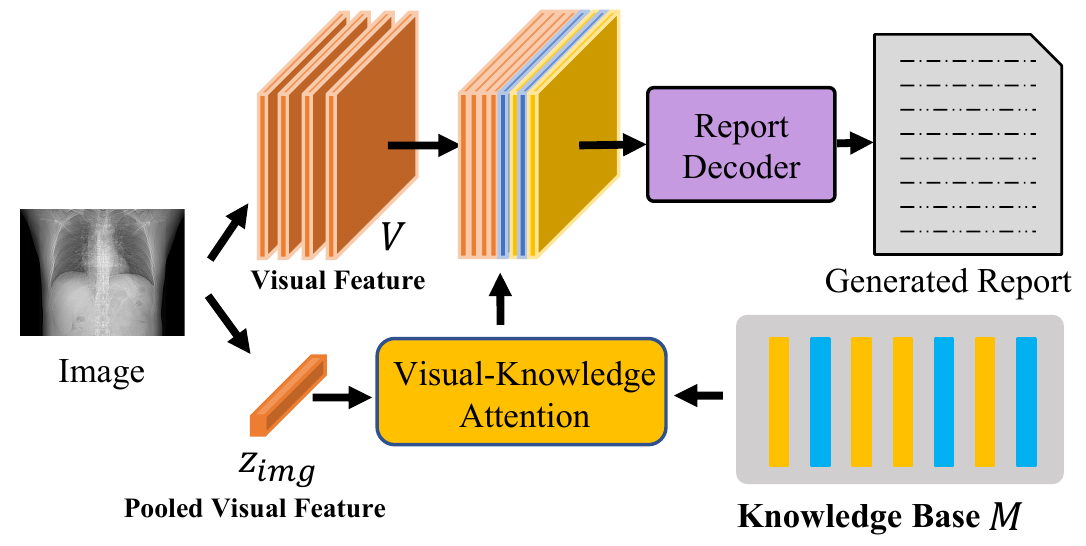}
  \caption{The architecture of the proposed model during inference stage. Here we obtain the visual features($V,z_{img}$) from the original image(Image) through a pre-trained visual encoder.}
  \label{fig:inference}
\end{figure}
\subsection{Knowledge Base Updating}
\label{sec:memorybank}
Different from image captioning, radiology report generation requires professional medical domain knowledge. The radiologists need to examine the diseases' degree, size, and location before writing reports. To distill the domain knowledge conceived in the pairs of radiology images, we introduce the memory mechanism. As revealed by recent studies ~\cite{Graves2016Hybrid, Wu2018,ChenImp2020}, the memory mechanism can provide prior knowledge to boost the generation model. Inspired by this, we initialize a memory module as a radiology knowledge base and propose an updating mechanism to learn from training data.

In our implementation, we use a matrix to initialize the knowledge base $\mathbf{M}_0 \in \mathbb{R}^{N_m \times D}$, where $N_{m}$ is the base size and $D$ is the dimension. In initialization, the diagonal element of $\mathbf{M}_0$ is set 1, and other elements are set to 0.
 
To model the relationship among the knowledge base, the textual features from the reference reports, and the visual features from input images, we use the multi-head attention (MHA)~\cite{Vaswani2017} mechanism. 
The MHA consists of $n$ parallel heads and each head is defined as a scaled dot-product attention:
\begin{align}
&\text{Att}_i(X,Y) = \text{softmax}\left(\frac{X\text{W}_i^\text{Q}(Y\text{W}_i^\text{K})^T}{\sqrt{{d}_{n}}}\right)Y\text{W}_i^\text{V}, \nonumber \\
& \text{MHA}(X,Y) = [\text{Att}_1(X,Y); \dots; \text{Att}_n(X,Y)]\text{W}^\text{O},
\end{align}
where $X \in \mathbb{R}^{l_x \times d}$ and $Y \in \mathbb{R}^{l_y \times d}$ denote the Query matrix and the Key/Value matrix, respectively; $\text{W}_i^\text{Q}, \text{W}_i^\text{K}, \text{W}_i^\text{V} \in \mathbb{R}^{d \times d_n}$ and $\text{W}^\text{O} \in \mathbb{R}^{d \times d}$ are learnable parameters, where ${d}_{n} = d / {n}$. $[\cdot;\cdot]$ stands for concatenation operation.

To update the knowledge base $\mathbf{M}_{t-1}$ at the training step $t-1$, we use the textual features extracted from the reference reports identifying the knowledge that is missing in $\mathbf{M}_{t-1}$.
\begin{align}
    \Delta \mathbf{M}_{t} = \text{MHA}(\mathbf{M}_{t-1}, \mathbf{T}),
    \label{eq:kv}
\end{align}
where $\Delta \mathbf{M}_{t}$ stands for the incremental knowledge acquired at the training step $t$ and $\mathbf{T}$ denotes the textual features acquired in Eq.(\ref{eqn:textfeature}).
By integrating the incremental knowledge, we can acquire the updated knowledge base at training step $t$.
\begin{align}
    \mathbf{M}_{t}  = \mathbf{M}_{t-1} + \text{Norm}(\Delta \mathbf{M}_{t}),
    \label{eq:add}
\end{align}
where \textit{Norm} refers to layer normalization to normalize the incremental knowledge.

Next, we acquire supporting knowledge regarding the current image by multi-head attention. 
\begin{align}
    \mathbf{M}^S = \text{MHA}(\mathbf{z}_{img}, \mathbf{M}_{t}),
\end{align}
where $\mathbf{M}^S$ refers to the supporting knowledge, $\mathbf{z}_{img}$ refers to the pooled visual feature acquired from Eq.(\ref{eq:zimg}).

The report generator automatically generates a medical report using visual features and the learned knowledge base. The Transformer-based model has proven effective in natural language processing~\cite{Devlin2019}. Thus, we employ a standard Transformer model as our report generator~\cite{Vaswani2017} which includes a stack of self-attention layers and masked self-attention layers.
The generator decodes the concatenation of visual features and the supporting knowledge to a sequence of hidden representation and follows the auto-regressive decoding process generating the radiology report $\mathbf{W}$ from the conditional distribution: 
\begin{align}
    p_{\theta}(\mathbf{W}|[\mathbf{V};\mathbf{M}^S]) = \prod_{t=1}^{N_W}p_{\theta}(w_t|w_{1:t-1},[\mathbf{V};\mathbf{M}^S]),
\end{align}
where $\theta$ refers to the model's parameters, $w_{1:t-1}$ denotes the generated words before time step $t$, and $\mathbf{V}$ refers to visual features acquired in Eq.(\ref{eqn:visualfeature}).

\begin{table*}[htbp]
  \centering
  \caption{The performances of our model compared with baselines on IU-Xray and MIMIC-CXR datasets. The best results are highlighted in bold. For the baselines marked by *, we cite the results reported in~\cite{Jing2019}. For the baselines marked by \#, we replicate the experiments by running their codes; the rest are cited from the original papers.} 
  \label{tab:baseline}
    \begin{tabular}{c|l|cccccc}
    \toprule
    Dataset & Model & BLEU-1 & BLEU-2 & BLEU-3 & BLEU-4 & CIDEr & ROUGE-L \\
    \midrule
    \multirow{10}[4]{*}{IU-Xray} 
          & S\&T*~\citep{vinyals2015show}  & 0.216  & 0.124  & 0.087  & 0.066  & 0.294  & 0.306  \\
          & SA\&T$^\#$~\citep{xu2015show} & 0.399  & 0.251  & 0.168  & 0.118  & 0.302  & 0.323  \\
          & AdaAtt*~\citep{lu2017knowing} & 0.220  & 0.127  & 0.089  & 0.068  & 0.295  & 0.308  \\
          & CMAS*~\citep{Jing2019Show}  & 0.464  & 0.301  & 0.210  & 0.154  & 0.275  & 0.362  \\
          & KERP~\citep{Li2019Knowledge}  & 0.482  & \textbf{0.325}  & 0.226  & 0.162  & 0.280  & 0.339  \\
          & R2Gen~\citep{Chen2020} & 0.470  & 0.304  & 0.219  & 0.165  & /     & 0.371  \\
          & CMCL~\citep{Liu2021Competence}  & 0.473  & 0.305  & 0.217  & 0.162  & /     & 0.378  \\
          & PPKED~\citep{Liu2021Exploring} & 0.483  & 0.315  & 0.224  & 0.168  & 0.351  & 0.376  \\
          & CA~\citep{Liu2021Contrastive}    & 0.492  & 0.314  & 0.222  & 0.169  & /     & 0.381  \\
\cmidrule{2-8}          
          & Our   & \textbf{0.497 } & {0.319 } & \textbf{0.230 } & \textbf{0.174 } & \textbf{0.407 } & \textbf{0.399 } \\
    \midrule
    \multirow{7}[4]{*}{MIMIC-CXR} 
          & S\&T$^\#$~\citep{vinyals2015show}  & 0.256  & 0.157  & 0.102  & 0.070  & 0.063  & 0.249  \\
          & SA\&T$^\#$~\citep{xu2015show} & 0.304  & 0.177  & 0.112  & 0.077  & 0.083  & 0.249  \\
          & AdaAtt$^\#$~\citep{lu2017knowing} & 0.311  & 0.178  & 0.111  & 0.075  & 0.084  & 0.246  \\
          & TopDown$^\#$~\citep{anderson2018bottom} & 0.280  & 0.169  & 0.108  & 0.074  & 0.073  & 0.250  \\
          & R2Gen~\citep{Chen2020} & 0.353 & 0.218 & 0.145 & 0.103 & /     & 0.277 \\
          & CMCL~\citep{Liu2021Competence}  & 0.334  & 0.217  & 0.140  & 0.097  & /     & 0.281  \\
          & PPKED~\citep{Liu2021Exploring} & 0.360  & 0.224  & 0.149  & 0.106  & /     & \textbf{0.284 } \\
          & CA~\citep{Liu2021Contrastive}    & 0.350  & 0.219  & 0.152  & 0.109  & /     & 0.283  \\
\cmidrule{2-8}          
          & Our   & \textbf{0.386 } & \textbf{0.237 } & \textbf{0.157 } & \textbf{0.111 } & \textbf{0.111 } & 0.274  \\
    \bottomrule
    \end{tabular}
\end{table*}

\subsection{Multi-Modal Alignment}
\label{sec:consistency}
We introduce the Multi-Modal Alignment module which aligns the visual, textual, and disease labels to guide the training of the proposed model.

\noindent\textbf{Textual-Textual Alignment}. Following the paradigm of natural language generation, our basic model maximizes the likelihood of generated reports by minimizing the cross-entropy loss. The model is optimized by the consistency between generated and reference reports, so we named it textual-textual (T-T) alignment.
\begin{align}
    \mathcal{L}_{T-T} = -\frac{1}{N_W}\sum_{i=0}^{N_W}\log p_\theta(\mathbf{W}|{[\mathbf{V}; \mathbf{M}^S]}),
\end{align}
where $\mathbf{W}$ is the generated sequence.

Besides the general textual-textual alignment, we propose two extra multi-modal alignments. The radiology report generation is a multi-modal task that aims to transform radiology images into reports. Since reports and disease labels describe the observations on the x-ray image, the semantic features among images, reports, and disease labels should be consistent. Following this intuition, we propose \textit{visual-textual(V-T) alignment} and \textit{visual-label(V-L) alignment} to encourage our model to be consistent among different modalities and guide the learning of the proposed model.

\noindent\textbf{Visual-Textual Alignment}
The visual-textual alignment module tries to make the features in radiology images and reference reports to be close. Firstly, we extract the pooled textual feature $\mathbf{z}_{txt}$ from the reference report and the pooled visual features $\mathbf{z}_{img}$ from the input image.
As illustrated in~\cite{Chauhan2020}, the triplet margin loss~\cite{Balntas2016}  performs well for joint training image and report features in the radiology image classification task. We adapt the image-to-text and text-to-image triplet margin loss losses which force the paired features closer than unpaired features in latent space. This helps us to model the bidirectional inter-relationship between image and report.
Given a target pair $\{\mathbf{z}_{img}, \mathbf{z}_{txt}\}$, we sample a negative pair(unpaired) $\{\mathbf{z}^{(n)}_{img}, \mathbf{z}^{(n)}_{txt}\}$ from the training set, and the visual-textual alignment is formulated as:
\begin{align}
    \mathcal{L}_{V-T} = &\max(0, \mu + d(\mathbf{z}_{img}, \mathbf{z}_{txt}) - d(\mathbf{z}_{img},  \mathbf{z}^{(n)}_{img})) \nonumber \\
    +&\max(0, \mu + d(\mathbf{z}_{txt}, \mathbf{z}_{img}) - d(\mathbf{z}_{txt},  \mathbf{z}^{(n)}_{txt})), \\
    \text{where } &d(\mathbf{z}_1, \mathbf{z}_2)= 1 - \frac{\mathbf{z}_1\cdot\mathbf{z}_2}{||\mathbf{z}_1||\cdot||\mathbf{z}_2||}. \label{eqn:dot}
\end{align}

\begin{align}
L= &\max(0, \mu + d(\mathbf{z}_{img}, \mathbf{z}_{txt}) - d(\mathbf{z}_{img},  \mathbf{z}^{(n)}_{img})).
\end{align}

In our implementation, $d(\cdot,\cdot)$ is a distance function to quantify the similarity of two embeddings, and $\mu$ is a margin parameter defined as Eq.(\ref{eqn:mu}), which is determined by the difference of disease labels between the target image and negative image.
\begin{align}
  \mu = \left\{\begin{matrix}
  0 & \mathbf{Y} = \mathbf{Y}^{(n)} \\
  \max(0.5, \frac{1}{N_L}\sum_i^{N_L}|y_{i} - y_{i}^{(n)}|) & \mathbf{Y} \ne \mathbf{Y}^{(n)}
\end{matrix}\right., \label{eqn:mu}
\end{align}
where $\mathbf{Y} \in \mathbb{R}^{N_L}$ and $\mathbf{Y}^{(n)} \in \mathbb{R}^{N_L}$ denote the disease labels of input image and negative sampled image, respectively.

\noindent\textbf{Visual-Label Alignment}.
Let $\mathbf{Y'} \in \mathbb{R}^{N_L}$ denote the predicted labels of current input image by the proposed model. Then the visual-label alignment is calculated as follows.

\begin{align}
    \mathbf{Y'} = \mathbf{z}_{img}W^L + b^L, \label{eqn:label}
\end{align}
where $W^L$ is a learn-able affine transformation, and $b^L$ is a bias.

Next, we adopt binary cross entropy loss to optimize the model on the consistency between visual and disease labels:
\begin{align}
    \mathcal{L}_{V-L} =  -\frac{1}{N_L}\sum_{i=0}^{N_L}&y_i\log\phi(y'_i),
\end{align}
where $y_i$ and $y'_i$ are ground-truth label and the predicted label in Eq.(\ref{eqn:label}), respectively, and $\phi(\cdot)$ denotes a sigmoid function. 

Finally, we optimize the proposed model with the textual-textual alignment $\mathcal{L}_{T-T}$, the visual-textual alignment $\mathcal{L}_{V-T}$, and the visual-label alignment $\mathcal{L}_{V-L}$. It is formulated as:
\begin{align}
    \mathcal{L} = \lambda_1\mathcal{L}_{T-T} + \lambda_2\mathcal{L}_{V-T} + \lambda_3\mathcal{L}_{V-L}, \label{eqn:loss}
\end{align}
where $\lambda_1, \lambda_2$, and $\lambda_3$ are coefficients to balance the three constraint terms.

\section{Experiment}
In this section, we evaluate the proposed model on MIMIC-CXR and IU-Xray datasets and conduct some ablation studies to analyze the performance of the proposed model and the effectiveness of each component.
\subsection{Dataset}

\textbf{MIMIC-CXR}. MIMIC-CXR~\cite{Johnson2019} is a large dataset that contains 377,110 chest x-ray images and 227,827 free-text radiology reports associated with these images for 65,379 patients. The dataset contains multi-view images, and we filter frontal and lateral view images following previous works~\cite{Chen2020}. The dataset is labeled for 14 common chest radiology observations derived from the free-text radiology reports by a label tool CheXpert~\cite{Irvin2019}.

\textbf{IU-Xray}. Indiana University Chest X-ray Collection~\cite{Demner-Fushman2016} is a public radiology examination dataset containing 3,955 radiology reports and 7,470 posterior-anterior/lateral view chest x-ray images. Each report consists of MeSH, indication, comparison, findings, and impression. For consistency, we employ the CheXpert to extract the labels as to the IU-Xray dataset.

The finding section in both datasets is used as the ground-truth reference report since it directly describes the observations on x-ray images. First, we filter out the reports without x-ray images or images missing the findings section. Then, each report is converted to lower case and filtered out the rare words with a minimum frequency of three, which results in 760/7866 unique words on IU-Xray and MIMIC-CXR datasets, respectively. The labels include 12 disease labels and 2 individual labels indicating "\textit{No finding}" and "\textit{Support device}". There is no official split of the IU-Xray dataset, so we follow the data split of previous SOTA work R2Gen~\cite{Chen2020} which splits the data into training, validation, and testing set using a ratio 7:1:2 without overlap in patients. For the MIMIC-CXR dataset, the official split is adopted. 

\subsection{Implementation Detail}
\label{sec:implement}
In our implementation, we employ the ResNet-101~\cite{He2016deep} backbone as our visual encoder pre-trained on ImageNet. 
The report encoder and generator are implemented by ourselves and trained from scratch. All hyper-parameters are selected by the performances on the validation set. The layer of the report encoder is set to 3. The dimension of the model is set to 512. All images are resized to 224$\times$224, and we use zero padding for each mini-batch report to keep the same length. We use the Adam optimizer with an initial learning rate of 5e-5 for fine-tuning the visual encoder. For other components, we use the Adam optimizer with an initial learning rate of 1e-4 and weight decay of 5e-5. We train the model with epochs of 50 and 30 for the IU-Xray and MIMIC-CXR datasets, respectively. The coefficients of multi-modal alignment $\lambda_1, \lambda_2$, and $\lambda_3$ are set to 1, 0.1, and 0.1. We evaluate the proposed model on the validation set and report the results on the testing set when the performance of the validation set achieves the best BLEU-4 score.

\subsection{Quantitative Results}
We adopt the widely used natural language generation (NLG) metrics and clinical efficacy metrics to evaluate the performance of the proposed model. The NLG metrics include BLEU-n~\cite{papineni2002bleu}, CIDEr~\cite{vedantam2015cider}, and ROUGE-L~\cite{lin2004rouge} score. The BLEU-n is used to measure the accuracy of the generated report. It is a widely adopted machine translation metric that analyzes the co-occurrences of n-grams between the generated sentences and ground truth: the higher degree of co-occurrences, the higher quality of the generated text. The ROUGE-L is measured similarly to BLEU-n, and the difference is that BLEU-n calculates the accuracy, and ROUGE-L calculates the recall. The CIDEr score evaluates whether the text generated by the model covers the key information in the ground truth. It regards each sentence as a document, calculates the cosine of the TF-IDF vector, and compares the similarity between the generated text and the ground truth text in the vector space. The higher of the value, the better the performance of the model. We compute these metrics using the MSCOCO caption evaluation tool~\footnote{https://github.com/tylin/coco-caption}. 

The clinical efficacy (CE) metrics are proposed in R2Gen~\cite{Chen2020}, including accuracy, precision, recall, and F1 scores of the disease labels between the reference and generated radiology reports. The providers of the MIMIC-CXR dataset use the rule-based CheXpert labeler to extract medical terminology from reports, and the CE metrics are calculated by using the CheXpert labeler to extract labels from the ground-truth and generated reports automatically. Since the IU-Xray dataset does not provide consistent labels, we only report CE metrics on the MIMIC-CXR dataset. The CE metrics can help to evaluate how well the generated reports describe abnormalities.

We compare our proposed model with general image captioning works, e.g., \textbf{S\&T}~\cite{vinyals2015show}, \textbf{SA\&T}~\cite{xu2015show}, \textbf{AdaAtt}~\cite{lu2017knowing}, and \textbf{TopDown}~\cite{anderson2018bottom}, and specific medical report generation works, e.g., \textbf{CMAS}~\cite{Jing2019}, \textbf{KERP}~\cite{Li2019Knowledge}, \textbf{R2Gen}~\cite{Chen2020}, \textbf{CMCL}~\citep{Liu2021Competence}, \textbf{CA}~\citep{Liu2021Contrastive}, and \textbf{PPKED}~\cite{Liu2021Exploring}. 

\begin{table}[htbp]
  \centering
  \caption{The results of clinical efficacy metrics on the MIMIC-CXR dataset. For the baselines marked by \#, we replicate the experiments by running their codes; the rest are cited from the original papers. }
    \begin{tabular}{c|cccc}
    \toprule
    Model & Accuracy   & Precision     & Recall     & F1 score \\
    \midrule
    S\&T$^\#$  & 0.423  & 0.084  & 0.066  & 0.072  \\
    SA\&T$^\#$ & 0.703  & 0.181  & 0.134  & 0.144  \\
    AdaAtt$^\#$ & 0.741  & 0.265  & 0.178  & 0.197  \\
    TopDown$^\#$ & 0.743  & 0.166  & 0.121  & 0.133  \\
    R2Gen & /     & 0.333  & 0.273  & 0.276  \\
    \midrule
    Ours   & \textbf{0.795 } & \textbf{0.420 } & \textbf{0.339 } & \textbf{0.352 } \\
    \bottomrule
    \end{tabular}%
  \label{tab:ce}%
\end{table}%

The NLG and CE results are shown in Table~\ref{tab:baseline} and Table~\ref{tab:ce}, respectively. Our proposed method achieves the state-of-the-art(SOTA) performance on almost all the metrics. Especially in terms of the CE metrics, our method remarkably outperforms all the previous models, leading to an increase of 26.1\% on precision,24.2\% on recall, and 27.5\% on the F1 score, which indicates that our model generates more accurate reports than others in the clinical perspective. Our method also improves the CIDEr score by a large margin, with an increase of 16.0\% on the IU-Xray dataset and 32.1\% on the MIMIC-CXR dataset. The higher CIDEr score indicates that our model can cover more key information in the generation. Besides, our higher BLEU-n and ROUGE scores indicate that our model can generate more cognitively fluent sentences. The results demonstrate that the proposed multi-modal alignment and learned knowledge base can better assist visual feature extraction during training and provide useful radiology knowledge.

\subsection{Ablation Study}
We conduct experiments to analyze the effectiveness of the multi-modal alignment mechanism, the impact of the knowledge base size, and the impact of the head size in knowledge base updating. 

\begin{table}[htbp]
  \centering
  \caption{The performance of models with different alignment methods on both the IU-Xray and MIMIC-CXR datasets. The V-T and V-L refer to visual-textual alignment and visual-label alignment, respectively. The first row of ``-'' refers to the model without any alignment.}
    \resizebox{0.45\textwidth}{!}{
    \begin{tabular}{c|c|ccc}
    \toprule
    Dataset & Alignment & BLUE-3 & BLUE-4 & ROUGE-L \\
    \midrule
    \multirow{4}[2]{*}{IU-Xray} & -     & 0.207  & 0.157  & 0.356  \\
          & V-L   & 0.215  & 0.163  & 0.371  \\
          & V-T   & 0.221  & 0.162  & 0.382  \\
          & V-T + V-L & \textbf{0.230 } & \textbf{0.174 } & \textbf{0.399 } \\
    \midrule
    \multirow{4}[2]{*}{MIMIC} & -     & 0.141  & 0.099  & 0.270  \\
          & V-L   & 0.149  & 0.105  & 0.271  \\
          & V-T   & 0.144  & 0.102  & \textbf{0.274 } \\
          & V-T + V-L & \textbf{0.157 } & \textbf{0.111 } & \textbf{0.274 } \\
    \bottomrule
    \end{tabular}%
    }
  \label{tab:component}%
\end{table}%

Table~\ref{tab:component} shows the performances of our model using different alignments on both datasets. The first three rows show that the model with visual-label alignment and the model with visual-textual alignment outperform the model without any alignment. In other words, our model benefits from each kind of alignment, which proves the effectiveness of the proposed two alignments. Furthermore, the model that combines both alignments achieves the best performance, which indicates the two alignments can assist each other and boost the quality of the generated reports. 

\begin{table}[htbp]
  \centering
  \caption{The performance of different number of heads in knowledge base updating. \textit{\#Head} refers to the number of head. The BL-n, CDr, and RG-L denote BLEU-n, CIDEr, and ROUGE-L scores, respectively.}
    \begin{tabular}{c|c|cccc}
    \toprule
    Dataset    & \#Head  & BL-3  & BL-4  & CDr   & RG-L \\
    \midrule
    \multicolumn{1}{c|}{\multirow{4}[2]{*}{IU-Xray}} & 1     & 0.205  & 0.149  & 0.289  & 0.372  \\
          & 2     & 0.216  & 0.158  & 0.379  & 0.379  \\
          & 4     & 0.225  & 0.164  & 0.371  & 0.379  \\
          & 8     & \textbf{0.230 } & \textbf{0.174 } & \textbf{0.407 } & \textbf{0.399 } \\
    \midrule
    \multicolumn{1}{c|}{\multirow{4}[2]{*}{MI\newline{}MIC}} & 1     & 0.131  & 0.090  & 0.109  & 0.259  \\
          & 2     & 0.144  & 0.101  & 0.103  & 0.271  \\
          & 4     & 0.149  & 0.105  & \textbf{0.120 } & 0.271  \\
          & 8     & \textbf{0.157 } & \textbf{0.111 } & 0.111  & \textbf{0.274 } \\
    \bottomrule
    \end{tabular}%
  \label{tab:head}%
\end{table}%

\begin{figure*}[htbp]
  \centering
  \includegraphics[width=0.98\linewidth]{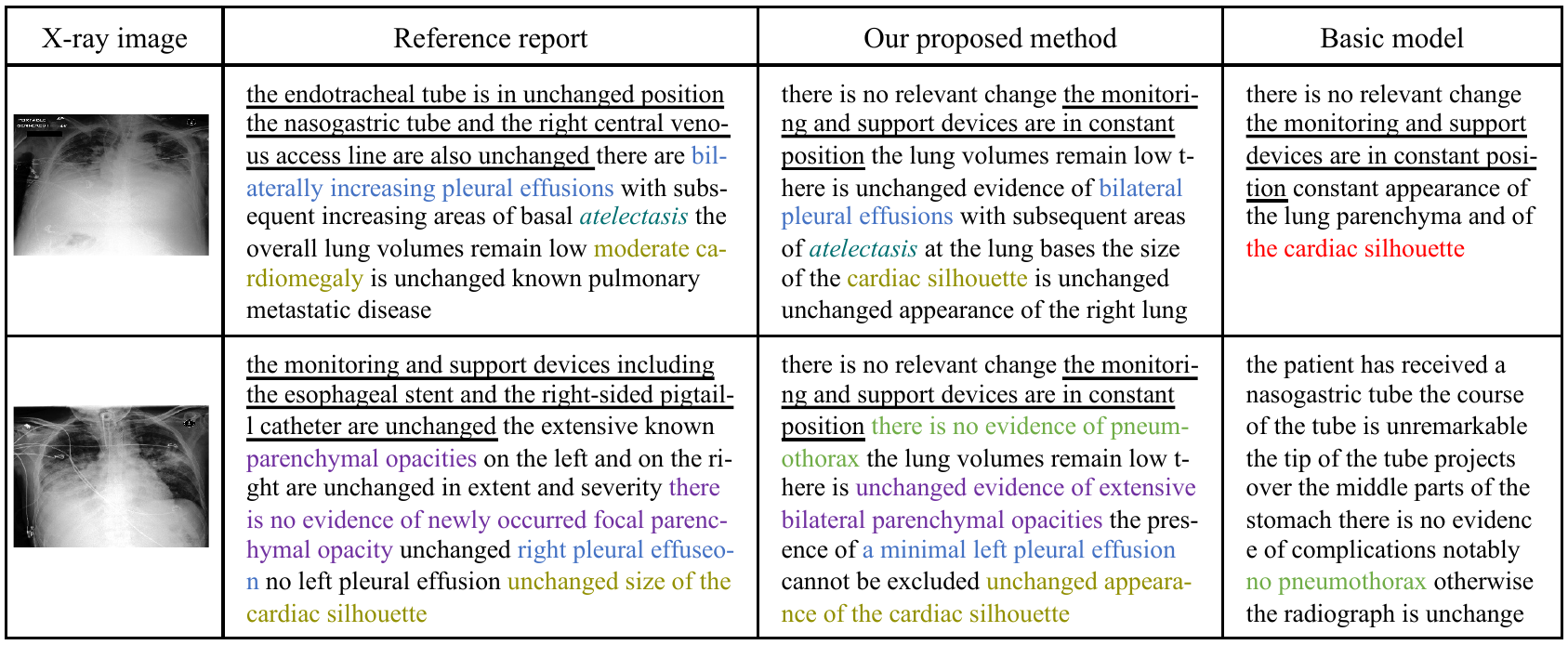}
  \caption{Visualization of two samples with reports from the MIMIC-CXR dataset. The same color highlights the descriptions of the same disease. The descriptions of the \textit{support device} are highlighted by underlining.}
  \label{fig:case}
\end{figure*}

To validate the impact of the different number of heads in knowledge base updating, we conduct experiments to evaluate the performance of models with different numbers of heads. We train four models with the number of heads as 1, 2, 4, and 8. As shown in Table~\ref{tab:head}, the number of heads increases and the model's performance improves consistently, showing the enhanced expressive ability to use the multi-head mechanism to learn the features from different sub-spaces. The results demonstrate the effectiveness of the proposed knowledge base updating with multi-head attention.

\begin{table}[htbp]
  \centering
  \caption{The performance of our model with different size of knowledge bases. The \textit{\#KB} denotes the size of knowledge base, where 0 refers to the model without knowledge base. The BL-n, CDr, and RG-L denote BLEU-n, CIDEr, and ROUGE-L scores, respectively.}
    \begin{tabular}{c|c|cccccc}
    \toprule
    Dataset    & \#KB    & BL-3  & BL-4  & CDr   & RG-L \\
    \midrule
    \multicolumn{1}{c|}{\multirow{5}[2]{*}{IU-Xray}} & 0     & 0.197  & 0.149  & 0.366  & 0.376  \\
          & 1     & 0.206  & 0.154  & 0.386  & 0.371  \\
          & 10    & 0.211  & 0.155  & 0.375  & 0.368  \\
          & 30    & \textbf{0.230 } & \textbf{0.174 } & 0.407  & \textbf{0.399 } \\
          & 60    & 0.227  & 0.173  & \textbf{0.416 } & 0.395  \\
    \midrule
    \multicolumn{1}{c|}{\multirow{5}[2]{*}{MIMIC}} & 0     & 0.126  & 0.091  & 0.104  & 0.269  \\
          & 1     & 0.138  & 0.092  & 0.109  & 0.267  \\
          & 10    & 0.148  & 0.104  & 0.106  & 0.270  \\
          & 30    & 0.152  & 0.107  & 0.108  & 0.272  \\
          & 60    & \textbf{0.157 } & \textbf{0.111 } & \textbf{0.111 } & \textbf{0.274 } \\
    \bottomrule
    \end{tabular}%
  \label{tab:memorysize}%
\end{table}%

The size of the knowledge base determines the capacity of the knowledge learning from reference reports. We conduct the experiments to verify the contribution of the knowledge base component.As shown in Table~\ref{tab:memorysize}, we train five models with sizes of 0, 1, 10, 30, and 60, where the first row(size=0) indicates model without the knowledge base. the performance of our model goes up as the size rises, but it achieves the best performance when the size is 30 on the IU-Xray dataset. The potential reason is that the IU-Xray dataset is relatively small, so the diversity is limited. As the size goes up, the redundant information stored may distract the model, leading to declined performance. The performance improves continuously on the MIMIC-CXR dataset, showing that the model benefits from the large knowledge base on a large dataset. Compared with the model without the knowledge base, the performance of our model is significantly improved.

\subsection{Qualitative Results}
\label{sec:quanlitative}

In Figure~\ref{fig:case}, we visualize two images with reports from the MIMIC-CXR dataset. Our model generates more accurate descriptions from the x-ray image than the basic model. For example, in the first row, our model accurately generates the descriptions of \textit{pleural effusions, atelectasis, and cardiomegaly}. In contrast, the basic model lacks the key descriptions of \textit{pleural effusions, atelectasis}. The benefit may be due to the guiding of the multi-modal alignment mechanism and the learned knowledge base, which helps the model better capture the subtle changes in the radiology images. Furthermore, as we can see, our model predicts the descriptions of the support device with sentence \textit{``the monitoring and support devices are in constant position''} on both generated reports, which has the same meaning as the underlined sentences in the ground truth reports. However, the basic model only correctly describes the first report's \textit{support device} while missing such information on the second one, which demonstrates the superiority of our model of using the knowledge base to learn the template representation of the information {\em support devices}.

\section{Conclusion}
We proposed a novel report generation model for report generation with a learned knowledge base to learn and store medical knowledge automatically and a multi-modal alignment mechanism that promotes the semantic alignment among images, reports, and
disease labels to guide the learning of visual features. We conducted experiments to validate the proposed model and demonstrated the effectiveness of each component on both the IU-Xray and MIMIC-CXR datasets. The results show that our model achieves state-of-the-art performances in terms of NLG and CE metrics on both datasets. The ablation studies show that the multi-modal alignments can effectively guide the learning of visual features. We also show that our model performs better when the number of heads in the knowledge base updating and the size of knowledge bases are relatively large. It is worth mentioning that our proposed knowledge base updating method can easily be applied to various datasets as it can automatically build the knowledge base or template base without manual labor. In the future, we will continuously apply our method to other report generation tasks.

\section{Acknowledgments}
This work was supported by the CCF-Tencent Open Fund and Natural Science Foundation of China under Grant 31900979.

\newpage
\bibliography{aaai22}


\end{document}